\documentclass[preprint,showpacs,preprintnumbers,amsmath,amssymb]{revtex4}

\usepackage{graphicx}
\usepackage{dcolumn}
\usepackage{bm}

\begin{document}

\preprint{APS/123-QED}
\title{Resistivity study of the pseudogap phase for (Hg,Re) - 1223 superconductors}
\date{\today}

\author{C. A. C. Passos}
\email{cacpassos@yahoo.com}
\affiliation{%
High Pressure Laboratory-Preslab, Departamento de F\'{\i}sica, Universidade Federal do Esp\'{\i}rito Santo, Av. Fernando Ferrari 514, Vit\'oria,29060-900 - ES, Brazil
}%
\author{M. T. D. Orlando}
\affiliation{%
High Pressure Laboratory-Preslab, Departamento de F\'{\i}sica, Universidade Federal do Esp\'{\i}rito Santo, Av. Fernando Ferrari 514, Vit\'oria,29060-900 - ES, Brazil
}%
\author{J. L. Passamai Jr}
\affiliation{%
High Pressure Laboratory-Preslab, Departamento de F\'{\i}sica, Universidade Federal do Esp\'{\i}rito Santo, Av. Fernando Ferrari 514, Vit\'oria,29060-900 - ES, Brazil
}%
\author{E. V. L.  de Mello}
\affiliation{Departamento de F\'{\i}sica, Universidade Federal Fluminense, Niter\'oi - RJ, 24210-340, Brazil
}%
\author{H. P. S. Correa}
\affiliation{Departamento de F\'{\i}sica, Universidade Federal do Mato Grosso do Sul, Brazil
}%
\author{L. G. Martinez}
\affiliation{Centro de Ci\^encia e Tecnologia de Materiais - CCTM, Instituto de Pesquisa Energ\'eticas e Nucleares - IPEN, Campus USP, S\~ao Paulo - SP, 05508-900, Brazil
}%
\begin{abstract}
The pseudogap phase above the critical temperature of high $T_{c}$ superconductors (HTSC) presents different energy scales and it is currently a matter of intense study. The complexity of the HTSC normal state requires very accurate measurements with the purpose of distinguishing different types of phenomena. Here we have performed systematically studies through electrical resistivity ($\rho$) measurements by several different current densities in order to obtain an optimal current for each sample. This approach allows to determine reliable values of the pseudogap temperature $T^{*}(n)$, the layer coupling temperature between the superconductor layers $T_{LD}(n)$, the fluctuation temperature $T_{scf}(n)$ and the critical temperature $T_{c}(n)$ as function of the doping $n$. The interpretation of these different temperature scales allows to characterize possible scenarios for the (Hg,Re) - 1223 normal state. This method, described in detail here, and used to derive the (Hg,Re)-1223 phase diagram is general and can be applied to any HTSC.
\end{abstract}


\maketitle

\section{Introduction}
One of the greatest puzzles of the condensed matter physics is to understand the superconducting fundamental interactions of HTSC and their many unconventional properties. Among these, the pseudogap region below $T^{*}$ and above $T_{c}$, the so-called pseudogap phase~\cite{tim} has attracted a lot of attention in order to determine its precise values as function of the doping level $n$ and, above all, to understand its nature and its relation to the superconducting phase. Therefore a large number of different techniques have been used to study the dependence of $T^{*}$ for many family of compounds \cite{tim,tal}. However the major problem is that the values of $T^{*}$ differ strongly for different techniques and even a given method may yield different values. In many cases, different temperatures or energy scales are identified in the normal phase \cite{mou}, what makes extremely difficult to precisely understand the nature of this phase. Thus, one can find in the literature some different phase diagram for HTSC. As concerns its nature, many theoretical explanations have been proposed but they can be roughly classified in two main proposals. One is based on the fluctuation  of Cooper pairs between $T_c$ and $T^{*}$  with a non-vanishing order parameter without phase coherence or long range order \cite{eme}. Other proposal is based on the existence of some other type of order which may compete with the superconducting order \cite{tal,lee,mel2,mel3}.

We attempt here to define a systematic approach to study the pseudogap phase. Among the many different techniques used to solve this problem \cite{tim,tal}, transport properties by electrical resistivity measurements have been considered one of the most useful one. At high temperatures, the resistivity ($\rho$) has a linear behavior with the temperature and $T^{*}$ is defined as the temperature in which $\rho(T)$ starts to decrease below such linearity. However, it is well known that there are considerable differences in the values of $T^{*}$ found in some published works \cite{fuk,mel1,wan,wuy,she}. It is very likely that the discrepancy at $T^{*}$ has its origin in the fact that there are many parameters, which can determine the accuracy of the resistivity measurements. As considering polycrystalline samples, these factors are: the morphology of the junctions, the cross section of the grains and stoichiometry in the grain \cite{passos}. However, the most important factor is the applied current density value and its influence on the resistivity measurements in polycrystal or single crystal. In order to obtain precise values of $T^{*}$, it is crucial to perform the measurements in the linear or low current regime but, if the current is too low, the values could be plagued by high noise . Therefore, the precision of resistivity measurements is an open question, that means: what would be the ideal current applied to a polycrystalline sample? A search in the literature shows that there is no consensus about what is the ideal value to be used at the four-point probe, and as will be discussed below, they differ by several orders of magnitudes in the literature.

In this paper, we outline a method in which the values of the voltage $V=V(I)$ and $ \rho(T)$ are measured by several values of current \textit{I} and temperature \textit{T}, in order to find a range of current density which is ideal to determine $T^{*}$. In simple terms, we search for the maximum current density value that is in the limit of linear response. So far as we know, there is not any published systematic analysis of this type despite its importance. Taking advantage of the precision of our data we also discuss the Lawrence-Doniach temperature criteria generalized by Klemn \cite{klemn} in the case of several superconducting layers in a periodicity length \cite{ramalho}. The analysis of the $T_{LD}$ is very interesting because it gives a feeling of the superconducting coupling among the layers. We have also used these data to discuss the thermodynamic fluctuations of the Cooper pairs through the phase fluctuation temperature $T_{scf}$ \cite{naq}, which is important to the discussion of the pseudogap scenario. This paper is organized as follows: in Sec. II we  discuss the experimental details of sample preparation, characterization, and resistivity measurements. In section III we describe our systematic study of the applied current linear regime. We discuss how to calculate $T^{*}$, $T_{LD}$ and $T_{scf}$. These temperatures together with $T_c$ provide a possibility to discuss the phase diagram. 
\section{Experimental Details}
\subsection{Superconductor synthesis}
The ceramic precursor preparation began with a mixture of $Ba_{2}Ca_{2}Cu_{3}O_{x}$ (99.0\% PRAXAIR) and $ReO_{2}$ (99.0\% Aldrich) in powder form with the molar relationship 1 : 0.18 \cite{mtdo}. These powders were homogenized in an agate mortar and pelletized with an applied uniaxial pressure of 0.5 GPa. The produced pellet was heated at $850\,^{\circ}\mathrm{C}$ in a flow of oxygen (99.5\% purity) for 15 h. The obtained precursor was crushed, homogenized and compacted again before being heated a second time at $920\,^{\circ}\mathrm{C}$ for 12 h in a flow of oxygen. The later procedure is repeated for seven more times. These thermal treatment processes provide a good homogenization of the rhenium atoms and to eliminate the carbonates remaining in the precursor sample \cite{lou} as discussed in more detail in a previous work \cite{mtdo}.

The precursor material was submitted to an annealing at $920\,^{\circ}\mathrm{C}$ for 24 h in a flow of gas mixture of argon (99.5\% purity) and oxygen (99.5\% purity) maintained at 1 bar. Three different ceramic precursors were prepared with distinct partial pressure of oxygen $PO_{2}$: 5\% of $O_{2}$ and 95\% of Ar (sample A), 10\% of $O_{2}$ and 90\% of Ar (sample B) and 15\% of $O_{2}$ and 85\% of Ar (sample C) \cite{mtdo,fdco}. 

Finally, the three precursors prepared with different $PO_{2}$ were blended with \textit{HgO} at molar relationship 1 : 0.82. They were also homogenized in an agate mortar and palletized with an uniaxial pressure of 1 GPa. The pellets with a typical dimensions 5~$\times$~5~$\times$~20 mm$^{3}$  were wrapped in a gold foil (99.999\%) and introduced in a quartz tube with 8 mm inner diameter. Furthermore, it was introduced together with each pellet a rod of quartz (7 mm diameter and 40 mm length). Each sample (A, B and C) wrapped with a gold foil has received an excess of Hg (l) in amalgam form. The ratio between the mercury mass and the gold mass was 0.045. Basing on the study of the quartz tube filling factor effect (\textit{ff}), it was used \textit{ff}~$\cong$~1.0~g.cm$^{-3}$ and \textit{ff}$_{Hg}$ $\cong$ 0.010 g.cm$^{-3}$ \cite{mtdo2}. The quartz tubes were sealed in a high vacuum of 3$\times10^{-6}$ torr. All procedures were taken place inside a glove box filled with argon gas. In order to improve the grain size growth, the annealing time was 72 h at $865\,^{\circ}\mathrm{C}$ \cite{passos}. Moreover, three sealed quartz tubes, each one with a sample inside, were installed together in the same place inside an isostatic pressure furnace.  The furnace was filled with 14 bar of Ar to avoid an explosion of any quartz tube. 

\subsection{Superconductor characterization}
\subsubsection{X-ray diffraction measurements}
X-ray diffraction (XRD) analysis with Rietveld refinement were done in A, B, and C samples with the purpose of completing Orlando \textit{et} \textit{al.} \cite{mtdo} study. The XRD measurements were performed using laboratory diffractometers models Rigaku Multiflex and D-MAX with $CuK\alpha$ radiation. The spectra were measured from $2^{\circ}$ up to $122^{\circ}$ with step size of $0.01^{\circ}$ and counting time varying from 15 to 25 seconds, using very narrow slits to limit the X-ray beam. The instrumental parameters were obtained from the refinement of standards $LaB_{6}$ and $Al_{2}O_{3}$ (NIST  standards reference materials) samples. Rietveld refinements \cite{Rietveld} were performed using the program GSAS \cite{GSAS} with the interface EXPGUI \cite{EXPGUI}.
\begin{table*}
\caption{Results of the Le Bail fifs and crystallite sizes obtained from the XRD profile breadths. The Hg,Re-1223 and Hg-1223 phases are labeled by phase 1 and phase 2, respectively.} \label{rietveld}
\begin{ruledtabular}
\begin{tabular}{llccc}
\hline
  &Parameter& Sample A & Sample B & Sample C \\ 
 \hline
         & \% (Hg,Re)-1223 & 61.4 & 68.7 & 50.3 \\
         & \% Hg-1223 & 26.1 & 24.7 & 40.8 \\
 \hline
 phase 1 ~& \textit{a} (\AA)~& 3.854516(14)  ~& 3.854120(12)   ~& 3.854382(16)\\
         ~& c (\AA) ~& 15.687440(40) ~& 15.688061(56)   ~& 15.689091(70) \\
          & \textit{l} (\AA) & $>$ 1000  & $>$ 1000  ~& $>$ 1000 \\ 
 \hline
 phase 2 ~& \textit{a} (\AA) ~& 3.854295(18)  ~& 3.853526(15)  ~& 3.854320(10)\\
         ~& c (\AA) ~& 15.698784(60)  ~& 15.701567(65)  ~& 15.692780(76) \\
         & \textit{l} (\AA) & 590 & 380 & 470\\ 
 \hline
         & $\chi^{2}$ & 1.465 & 1.882 & 1.496 \\ 
         & Rwp (\%) & 3.83  & 3.03 & 3.70 \\
\hline
\end{tabular}
\end{ruledtabular}
\end{table*}

A typical Rietveld plot is shown in figure 1. For each XRD pattern, the better spectrum fit was obtained including an extra Hg-1223 phase additionally to the main (Hg,Re)-1223 phase, as compared to our previous work \cite{mtdo}. All refinements have considered the following phases: (Hg,Re)-1223 (rich at oxygen) and Hg-1223 (poor at oxygen), $HgCaO_{2}$, $BaCO_{3}$, $CaCuO_{2}$ and $BaCuO_{2}$ \cite{mtdo5,luiz}. The main (Hg,Re)-1223 and Hg-1223 phases, their fitted parameters, and good-of-fitness are shown in the table \ref{rietveld} \cite{luiz}. The existence of two superconducting phases was first detected by anomalous X-ray diffraction carried out at 8950 eV and 10600~eV at Brazilian Synchrotron Light Source (LNLS) - Campinas - Brazil \cite{furlan,mtdo6}. Moreover, it was confirmed by anomalous X-ray diffraction that Re distribution on the Hg-O plane did not produce a super cell in any sample (A, B and C) \cite{luiz,mtdo6}.
\begin{figure}[b]
\begin{center}
\includegraphics[width=12cm,height=9cm]{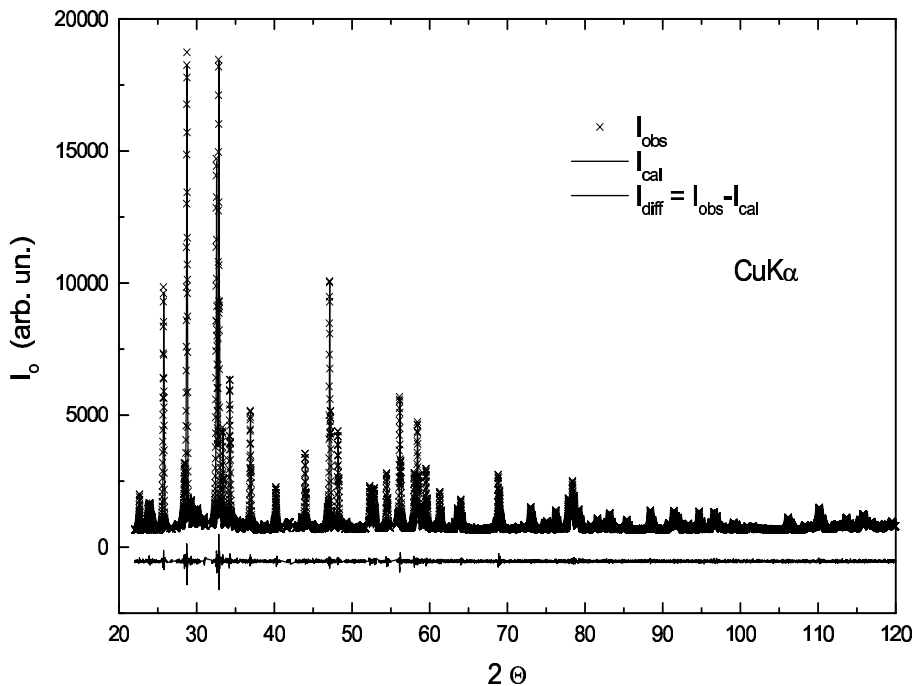}
\caption{Rietveld plot of the sample A (underdoped). The spectrum was plotted in a range from 20$^{\circ}$ up to 120$^{\circ}$. All XRD measurements were done with CuK$\alpha$ radiation.}
\end{center}
\end{figure}

The main (Hg,Re)-1223 phase was very crystalline as considering small broadening of their peaks. Besides, the crystallite average sizes were determined from the pseudovoigh profile coefficients of Le Bail fitting, using the formalism of Steffens \cite{GSAS,steffens}, Thompson approach \cite{thom}, and the Finger asymmetry correction \cite{finger}.

For all  samples the estimated crystallite size to the (Hg,Re)-1223 phase was larger than the range measurable by this method ($l \sim$ 1000 \AA). This indicates that during the final step of the synthesis occurs a strong growth of crystallites. On the other hand, the extra Hg-1223 phase has smaller crystallite as shown table \ref{rietveld}. Both phases did not present micro strains.

\subsubsection{SEM and EDS analysis}
As described in ref. \cite{passos}, the precursor annealing  influenced the oxygen partial pressure inside the sealed quartz tube. For the phase diagram region $PO_{2} < 0.2$ bar, the effect of the $PO_{2}$ pressure on the junction crystal size has been analyzed since 2000. With this aim in minds, we have obtained  Scanning Electron Microscopy (SEM) images. Using the image of sample B (see figure 2), a histogram of the grain-boundary size was done \cite{passos}. This procedure was also used for sample A and sample C. From these SEM images, the average junction sizes $\left\langle L\right\rangle$ were determined \cite{passos1} and are shown in table \ref{eds}. In addition, Energy Dispersion X-ray Spectra (EDS) analysis was done. These measurements have indicated the stoichiometry of the Hg, Re, Ba, Ca, and Cu elements present in the three samples. 
\begin{table}[!h]
\begin{ruledtabular}
\caption{\label{eds}Sample composition obtained by EDS measurements. The value \textit{$\left\langle L\right\rangle$} is the average junction size of the grain carried out by a SEM image analysis.} 
\begin{tabular}{ccc}
\hline
sample & Grain & $\left\langle L\right\rangle$ ($\mu$m) \\
\hline
A & $Hg_{0.83}Re_{0.17}Ba_{1.98}Ca_{2.01}Cu_{2.98}O_{8+\delta}$ &  2.1 \\
B & $Hg_{0.80}Re_{0.20}Ba_{1.99}Ca_{2.00}Cu_{2.98}O_{8+\delta}$ &  2.7 \\ 
C & $Hg_{0.79}Re_{0.21}Ba_{2.03}Ca_{1.98}Cu_{2.99}O_{8+\delta}$ &  2.4 \\
\hline
\end{tabular}
\end{ruledtabular}
\end{table}
It was shown in the micrographs a gradient in the content of \textit{Re} from the center to the boundary of the particle \cite{passos}. In our point of view, the Hg-1223 phase is preferentially formed in the periphery of the grains as a shell in which the crystallites were smaller than the center. Summarizing, the samples have similar morphology of the grains, average junction sizes, and same junction type (superconductor - insulate - superconductor), as reported in our recent work \cite{passos1}.
\begin{figure}[h]
\begin{center}
\includegraphics[width=12cm,height=11cm]{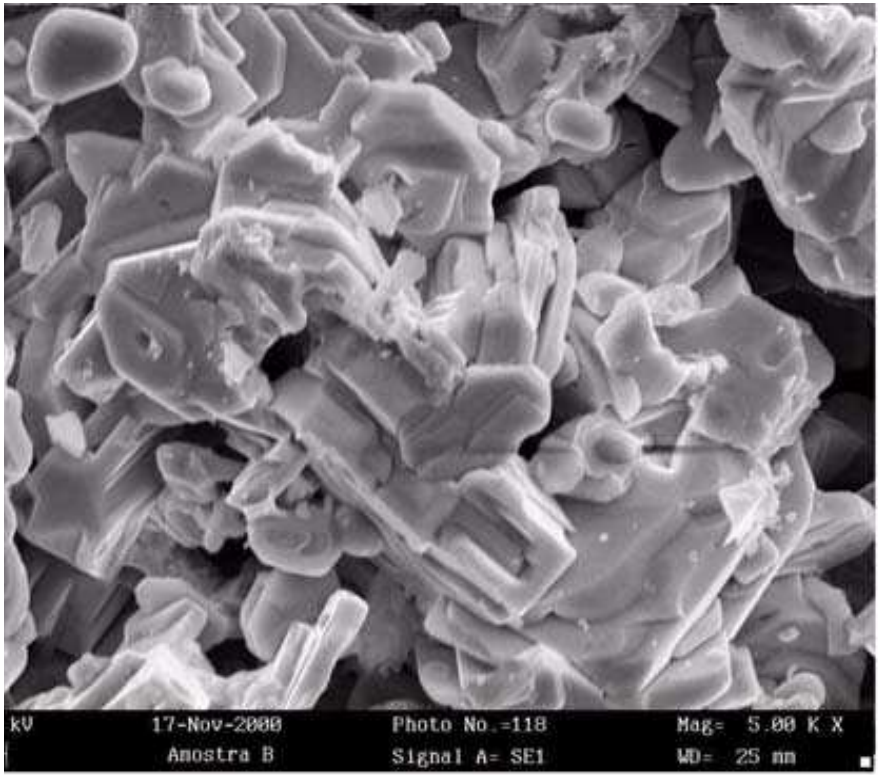}
\caption{SEM image performed on sample B (optimal doped). It can observe that the randomly oriented grain array, which is typical of a polycrystalline compound.}
\end{center}
\end{figure}
\subsubsection{AC susceptibility measurement}
The intergrain region of samples  were investigated by ac magnetic susceptibility ($\chi_{ac}$) using these samples in pellet form. Figure 3 shows the $\chi'_{ac}$ and $\chi''_{ac}$ components under distinct magnetic field (6, 16, 160 A/m) for sample B. In general, the out-of-phase component $\chi''_{ac}$ displays two peaks at distinct temperatures. The first is small and located close to $T_{c}$ and it is related to the intragrain-intrinsic-superconducting transition, which represents the statistical average bulk properties inside each grain of ceramic. The second peak of $\chi''_{ac}$ appears at lower temperature than the first and its shape depends on the characteristic intergrain connectivity (weak link region) of the grains in the superconductor ceramic sample \cite{sin1}. The in-phase component $\chi'_{ac}$ of the ac susceptibility presented two transitions towards lower diamagnetic screening. 
\begin{figure}[b]
\includegraphics[angle=90,width=12cm,height=9cm]{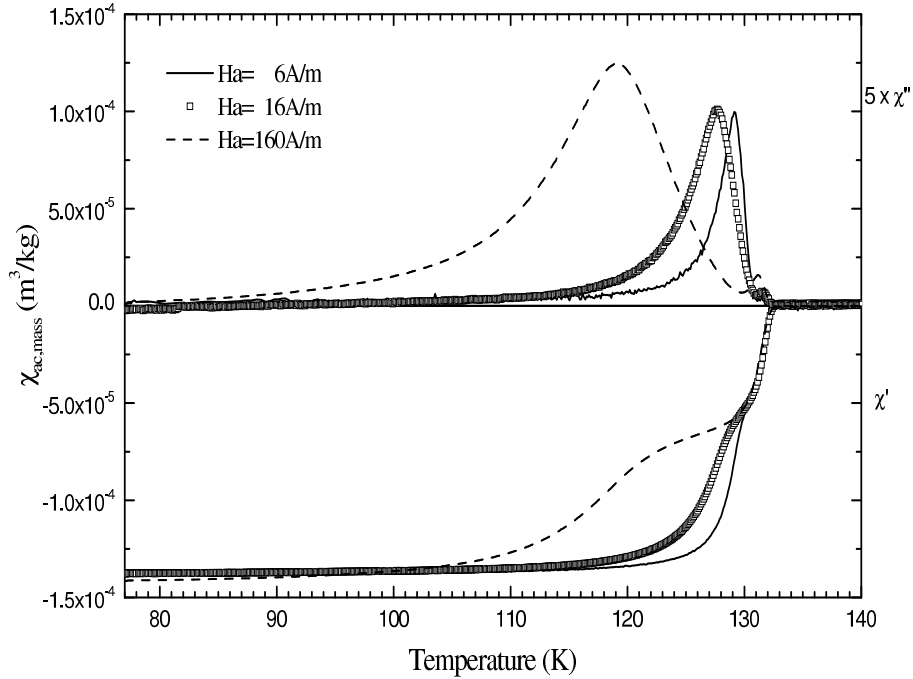}
\caption{AC magnetic susceptibility measurements presenting $\chi$' real part and $\chi$" imaginary part for sample B in the pellet form. The measurements were performed under distinct magnetic field at $\nu$ =43Hz.}
\end{figure}
The $T_{c}$ criterion for the (Hg,Re)-1223 phase was defined as the point where the $\chi'_{ac}$ signal is twofold the average noise value, which was used to measure before superconductor transition \cite{mtdo}. All details of the measurement procedures were reported elsewhere \cite{passos}. The onset critical temperatures $T_{c}$ for (Hg,Re)-1223 phase were (132.6 $\pm$ 0.2) K, (133.2~$\pm$~0.2) K and (132.7 $\pm$ 0.2) K for sample A, B, and C, respectively. The $\chi_{ac}$ measurements were repeated on the same samples in powder form (38$ \mu$m), and it was also yielded the same~$T_{c}$~\cite{passos}. 

For Hg-1223 phase, the second transition temperature was taken at split of $\chi'_{ac}$ signal, when different magnetic field amplitudes ($H_{a}$=6, 16, 160 A/m) and $\nu$ = 43 Hz were applied.  The second transition temperature values found in these conditions were (127.0 $\pm$ 0.2) K, (129.5 $\pm$ 0.2) K, and (127.0 $\pm$ 0.2) K. As intergrain morphologies are similar, these distinct temperatures are attributed to presence of the Hg-1223 phase in grain boundary and they are correlated with a variation of oxygen content. The exam of lattice parameters for the Hg-1223 phase (table \ref{rietveld}) reveals that this phase is underdoped and the expected $T_{c}$ would be (120 $\pm$ 3) K, (124 $\pm$ 3) K, and (120~$\pm$~3) K for samples A, B, and C, respectively, as considering the ref. \cite{paran}. Therefore, from room temperature down to 130 K the Hg-1223 phase is in normal state (non-superconductor).

The results suggest that the (Hg,Re)-1223 phase have similar oxygen contents, that is, alike physical properties for the samples were expected. Our conclusion is that the rhenium doping is a main oxygen fixing mechanism in (Hg,Re)-1223 compound. The Re atom provides extra oxygen atoms in the $HgO_{\delta}$ planes \cite{ochmai,mtdo}. For the Hg-1223 (without rhenium), the oxygen is found at the (1/2,1/2,0) are responsible for the doping variation since they are loosely bond to the Hg atoms. This mechanism leads to easy intercalation or removal of the oxygen during the synthesis. However, in the (Hg,Re)-1223 phase, there is a stronger Re-O bond, which has presented an oxygen at (0.33,0.33,0) \cite{mtdo}. The Re atom have also added or removed an extra oxygen in the crystallographic site (1,1/2,0) or (1/2,1,0). Therefore, the oxygen in the Re-O bond present in 20\% of the sample would be unlike the oxygen in the Hg-O bond and it may not be removed with a lower oxygen partial pressure present in the synthesis process. 

Sin \textit{et} \textit{al.} \cite{asin} have shown the phase diagram of (Hg,Re)-1223 as a function of the oxygen partial pressure ($PO_{2}$) of the precursor. For precursor prepared with $PO_{2} \leq$ 0.2 bar, it was found a high (Hg,Re)-1223 phase content and a slightly $T_{c}$ parabolic variation.  On the other hand, it was shown in our previous paper \cite{passos,mtdo} that the $T_{c}$ value is not enough parameter to define the oxygen contents in (Hg,Re)-1223 phase. 

It was already observed that the ac susceptibility measurement under external hydrostatic pressure is an important tool to confirm a small difference in oxygen doping. The samples in powder form have presented distinct $dT_{c}/dP$ values (8 $\pm$ 1), (1.9 $\pm$ 0.2) and (-1.6 $\pm$ 1) K/GPa, which was associated with under doped, optimal doped and over doped oxygen contents, respectively \cite{passos}. In recently work \cite{mtdo}, we have reported thermopower measurements that confirmed and determined the oxygen content in each sample.

\subsubsection{Resistivity setup measurements}
The dc electrical resistance of the samples was measured using the four-point probe method. The samples were cut in slab form with dimensions of 1.2 x 1.0 x 7.0 mm$^{3}$ and they were fixed on a sapphire sample holder by using GE varnish. The four contacts with low electric resistance (5 $\pm$ 1 $\Omega$) were attached to the samples with silver paint. A KEITHLEY 228A Current Source applied currents form 0.4 mA up to 10 mA at a fixed temperature, and the  corresponding voltage values were obtained using a KEITHLEY 182 sensitive digital voltmeter. 

The \textit{I-V} curves were measured reversing the current the direction during measurement in order to avoid contact resistance influence. The temperature was measured by a copper-constantan thermocouple attached to the sapphire and linked to the HP 34401A multimeter. A PC computer by IEEE-488 interface recorded all data. 
\begin{figure}[t]
\begin{center}
\includegraphics[angle=90,width=12cm,height=9cm]{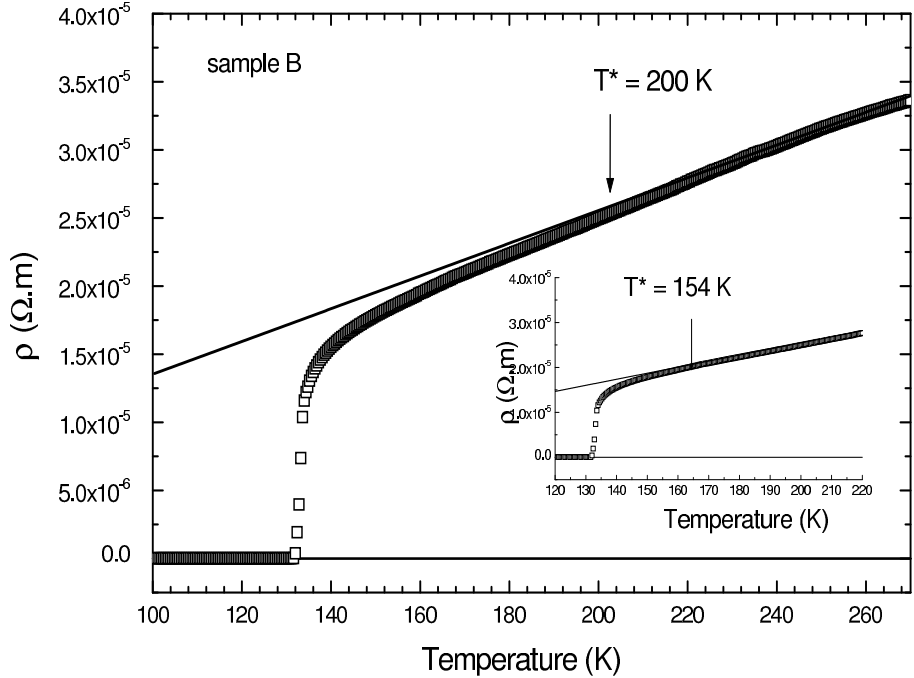}
\caption{The temperature dependence of electrical resistivity of the (Hg,Re)-1223 sample for $J$= 1.5 A/cm$^{2}$. The inset shows the $T^{*}$ determine from fitting up to 200K using $J$= 0.7 A/cm$^{2}$.}
\end{center}
\end{figure}
\section{Electrical resistivity analysis}
\subsection{Pseudogap temperature}

A search in the literature reveals that there is no consensus about what is the ideal current which must be used at the four-point probe, however it is crucial to be in the linear regime in order to calculate $T^{*}$. For instance, in the case of thin films, Qiu \textit{et} \textit{al.} \cite{qiu} have applied a current of 0.01 mA ($J \simeq$ 1.5 A/cm$^{2}$). Wuyts \textit{et} \textit{al.} \cite{wuy} have measured the temperature dependence of the resistivity with a current density $J \leq 10^{2}$ A/cm$^{2}$. For polycrystalline samples, Tristan Jover \textit{et} \textit{al.} \cite{tri} have used a current of 1.8 mA, while Batista-Leyva \textit{et} \textit{al.} \cite{bat} have used 0.35 mA. Gonz\'alez \textit{et} \textit{al.} \cite{go} have measured the resistivity within the linear response regime with a current density of 0.07 A/cm$^{2}$ applied to the $Hg_{0.82}Re_{0.18}Ba_{2}Ca_{2}Cu_{3}O_{8+d}$ sample or (Hg,Re)-1223 sample. In addition, Palstra \textit{et} \textit{al.} \cite{pal} have checked the linearity of the \textit{I-V} curves for currents between 0.1 and 100 mA in single crystals. In this ref. \cite{pal}, they have found that deviations from linearity start above 30 mA. Therefore, they have chosen a measuring current well below this value at 10 mA ($J \simeq$ 4.5 x  A/cm$^{2}$). As a consequence, although $T_{c}$ is very robust to these kind of current variations, it can occur an apparent discrepancy to define pseudogap temperature $T^{*}$. According to Tallon \textit{et} \textit{al.} \cite{tal}, when the resistivity measurement is plotted to 300~K, $T^{*}$ is equal to 195 K, which is defined as the temperature which $\rho(T)$ goes under the linear regime as the temperature goes down. However, when the same resistivity data are plotted up to 600 K,  a visual inspection yields $T^{*}$ at 320 K. 

As an example we show in figure 4 one of our typical resistivity measurement as function of the temperature for $J=$1.5A/cm$^{2}$. As one can see through the inset, if we use a value of $J=$0.7A/cm$^{2}$, the value of $T^{*}$ changes drastically. There are two factors responsible for such discrepancy: one is the pure visual analysis is subject of great uncertainties and must be replaced by derivative analysis. Other is the different values of $J$. As we show below  in figure 5, $J=$1.5A/cm$^{2}$ is already in the non-linear regime, and as a consequence, the value of $T^{*}$ is too high.

In general, the dependence of average voltage on the current can be written as the following expansion
\begin{equation}
V = R_{1}I + R_{2}I^{2} + \ldots,
\end{equation}
where $R_{1}=R_{1}(T)$ and $R_{2}=R_{2}(T)$. 

In figure 5 we plot the curves $V=V(I)$ for only one of our samples since the curves for the others have the same features. This figure shows  that, for $T=170K$, $V=V(I)$ is a linear function up to $J=4.2$ A/cm$^{2}$. However, for lower temperatures, non-linear effects develop in the $V(I)$ due to a non-vanishing value of $R_{2}$. As displayed in figure 5, for $T$=145K we have found that non-linear behavior appears above $J > 1.04$ A/cm$^{2}$. Thus, with the purpose of obtaining the high temperature linear behavior and the value of $T^{*}$ accurately,  one has to use $J \leq 1$ A/cm$^{2}$ for sample B. Accordingly, the same analysis for sample A yields the maximum linear current $J = 1.05$ A/cm$^{2}$ and for sample C, $J = 1.00$ A/cm$^{2}$. Thus, in what follows, we have taken the data with $J=1.0$ A/cm$^{2}$ for our three samples. 
\begin{figure}
\includegraphics[angle=90,width=11cm,height=7cm]{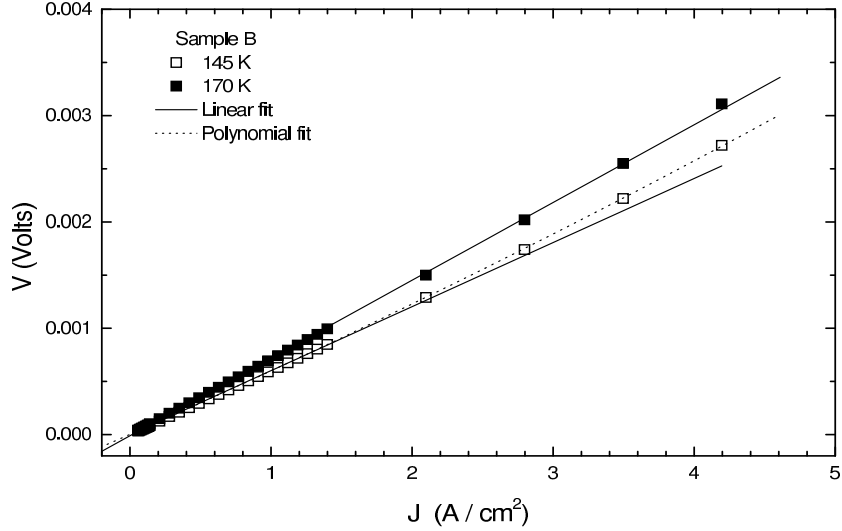}
\caption{Isotherms of the for sample B. The symbol ($\square$) denotes $V \times I$ curve  at T=145K and the symbol ($\blacksquare$) denotes $V \times I$ curve at T = 170K. The straight line is a linear regression fit and the dash line represents a polynomial (second rank) fit.}
\end{figure}

Taking into account the above optimum density current value, which assures us that the systems are in the linear regime ($R_{2}=0$) for $T>T_c$, we can write that
\begin{equation}
R_{1}(T)=R_{0} + \frac{\partial R}{\partial T}(T-T_{0}) + \frac{\partial^{2} R}{\partial T^{2}}(T-T_{0})^{2} + ...,
\end{equation}
where $T_0$ is any chosen temperature in the range of $T_c$ and what shows that the resistivity is linear with the temperature, whenever $\partial ^{2}R/\partial T^{2}=0$ and $\partial R / \partial T$ is independent of the temperature. Therefore, we can determine the values of $T^{*}(n)$ analyzing the first ($\partial \rho/ \partial T$) and second ($\partial ^{2}\rho/ \partial T ^{2}$) derivatives of the resistivity with respect to the temperature. The study of the regions, where the second derivative vanishes or change sign, has been used recently by Ando \textit{et} \textit{al.} \cite{and} to investigate the pseudogap phase of many compounds. Furthermore, Naqib \textit{et} \textit{al.} \cite{naq} have estimated $T^{*}$ above and below $T_{c}$ (by the use of Zn impurities) using the same method. They verified that for nearly identical values of number of holes, both sintered and high oriented thin film of $Y_{1-x}Ca_{x}Ba_{2}(Cu_{1-y}Zn_{y})_{3}O_{7-\delta}$ have the same values of $T^{*}$. The only difference between polycrystalline sample and thin film is the residual resistivity value. The polycrystal has a residual resistivity due to percolative effect and great contributions from the grain boundaries. 

As mentioned, the resistivity must be analyzed at low density of current to avoid the non-linear regime. On the other hand, lower values of $J$ are susceptible to high resistivity oscillations and do not allow an accurate estimation on $T^{*}$. Figures 6a, 6b, 6c show $(d\rho/dT)/(d\rho/dT)_{T = 170K}$ and $d^{2}\rho/dT^{2}$ for $J = $1 A/cm$^{2}$.  In these curves one can see that $(d\rho/dT)/(d\rho/dT)_{T = 170K}$  varies as the temperature is reduced down to $T_{c}$ but converges to a constant value as the temperature increases.  

The graphical analysis yielded $T^{*}$ = (160 $\pm$ 2) K for sample A, $T^{*}$ = (154 $\pm$ 2) K for sample B and $T^{*}$ = (151 $\pm$ 2) K for sample C. The uncertainties were estimated in the interval where the $(d\rho/dT)/(d\rho/dT)_{170K}$ curves start to deviate from the background and in the range of temperature which $d^{2}\rho/dT^{2}$ vanishes. 
\begin{figure}[b]
\includegraphics[angle=90,width=8cm,height=6cm]{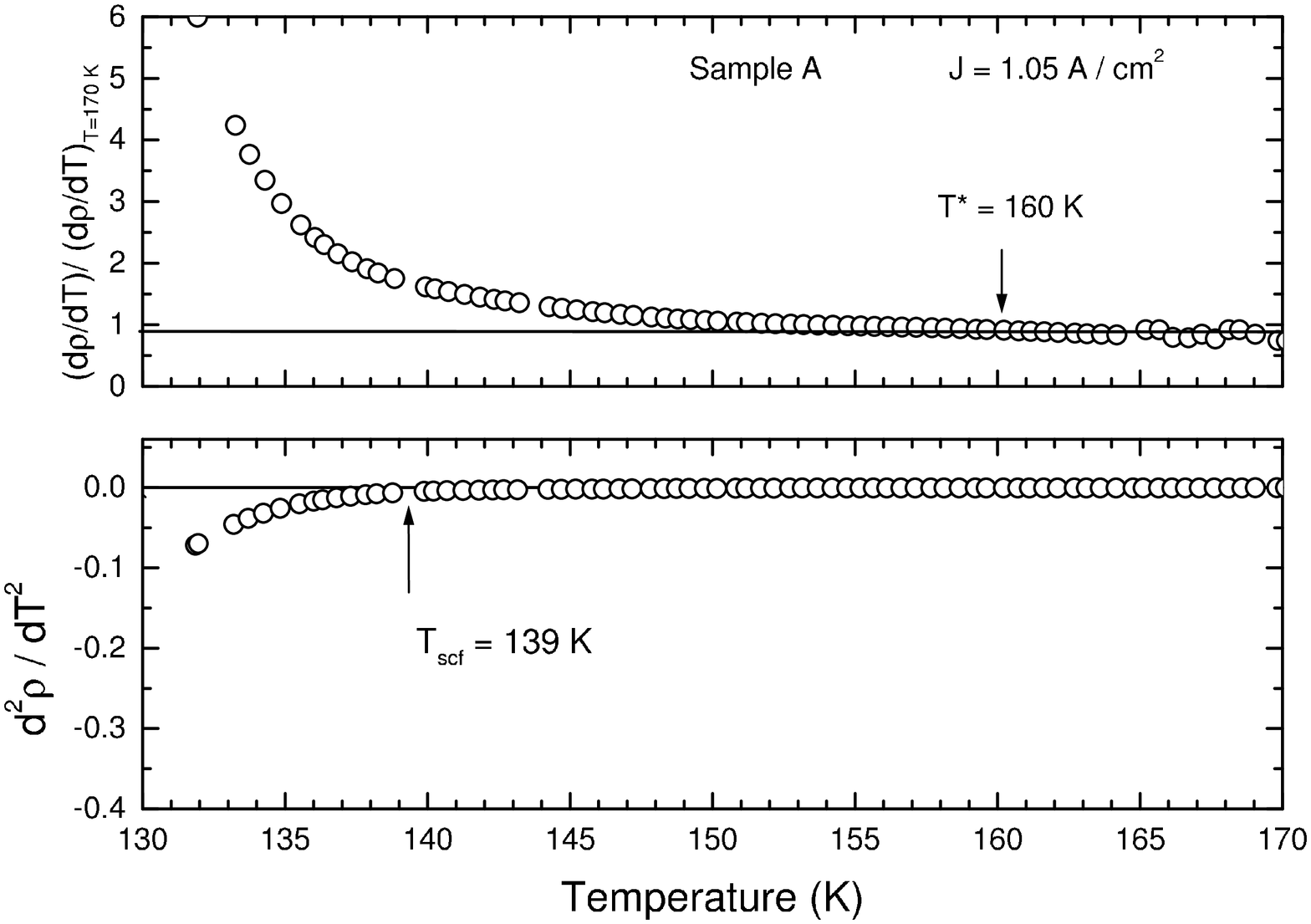}
\end{figure}
\begin{figure}
\includegraphics[angle=90,width=8cm,height=6cm]{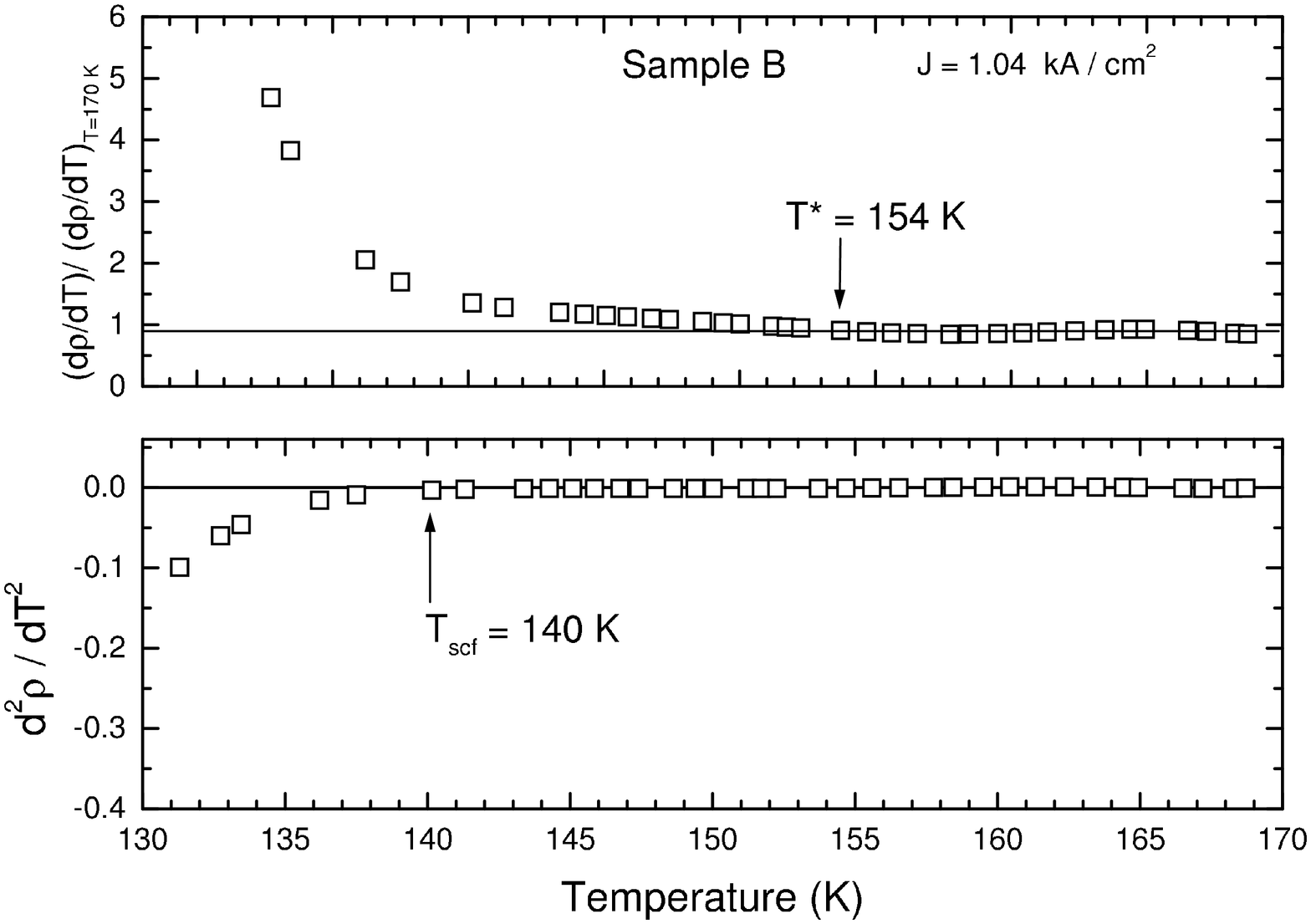}
\end{figure}
\begin{figure}
\includegraphics[angle=90,width=8cm,height=6cm]{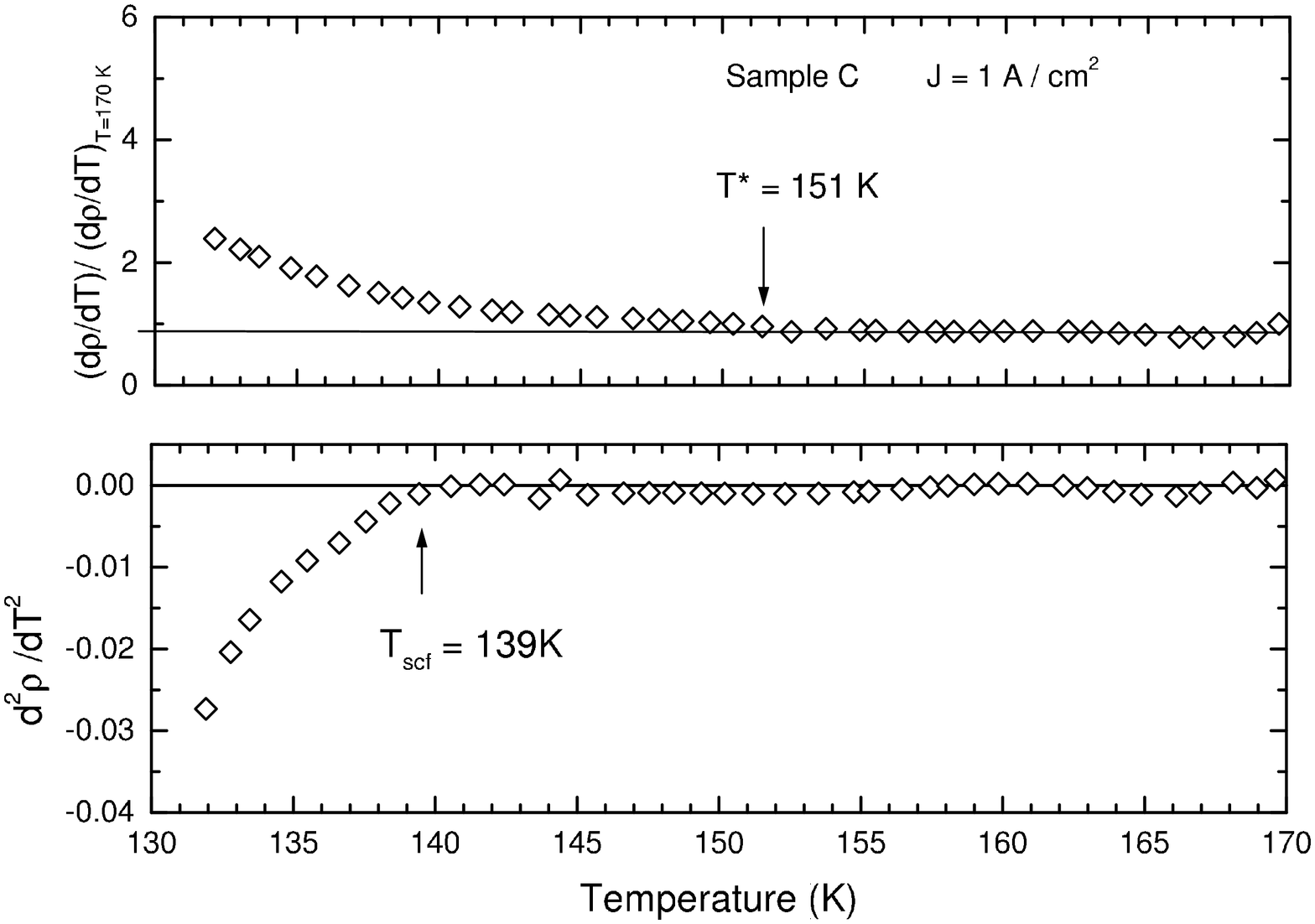}
\caption{The first and second derivative of resistance with respect to the temperature for (a) sample A , (b) sample B and (c) sample C. It was applied $J = $1 A/cm$^{2}$. $T_{scf}$ was defined from $d^{2}\rho/dT^{2}$ as the temperature at which strong and downturn in $\rho(T)$ becomes evident near $T_{c}$.}
\end{figure}
\newpage

\subsection{Superconducting Fluctuations}
As exposed in the introduction, one of the main proposal to the pseudogap phase is the existence of superconducting fluctuations without phase coherence \cite{eme,lee}. According this scenario, the HTSC exhibit complex behavior, which is related to thermodynamic fluctuations of the superconducting order parameter. These fluctuations affect the electrical resistivity characteristics in normal phase. 

For the polycrystalline samples case there are two models which can give a picture of fluctuations in intergrain and intragrain regions. The first model proposed by Aslamasov and Larkin \cite{asl} is associated with fluctuations in intergrain and  intragrain region, however the second developed by Lawrence and Doniach \cite{law} can be applied only for description of fluctuations into intragrain region of a layered superconductor.
\subsubsection{Aslamasov-Larkin model}
The thermodynamic fluctuations near the transition were first studied by Ginzburg \cite{gin}, and these effects in type I superconductor were shown to be negligible in 1960. However, the Aslamasov and Larkin report \cite{asl} have considered the effects of the superconducting fluctuations on the conductivity or paraconductivity to be non-negligible. Furthermore, the fluctuations are enhanced for sufficiently dirty films and whisker crystals \cite{asl}. Recently Naqib \textit{et} \textit{al.} \cite{naq} calculated the temperature where such fluctuations set in $T_{scf}$ and concluded that it is distinct of $T^{*}$, because they respond differently to an applied magnetic field. It is clear that the presence of Cooper pairs will affect the electrical resistivity. Therefore, following Naqib and co-workers, we used the same set of resistivity data to determine $T_{scf}$ and to estimate $T^{*}$ using the onset of vanishing of $d^{2}\rho/dT^{2}$ at a finite $d\rho/dT$ \cite{naq}. The results are displayed in table \ref{fluct}. The $T_{scf}$ values have a similar behavior of $T_{c}$, which presented a different trend than that of $T^{*}$ in agreement to Naqib \textit{et} \textit{al.} \cite{naq}.
\begin{table}[h]
\caption{\label{fluct} Comparison of the critical temperature, fluctuation conductivity and pseudogap temperature.}
\begin{ruledtabular}
\begin{tabular}{cccc}
\hline
Sample &  $T_{c}$(K) & $T_{scf}$(K)&  $T^{*}$(K)
 \\
\hline
A & 132.6 $\pm$ 0.2 & 139 $\pm$ 1  & 160 $\pm$ 2  \\
B & 133.2 $\pm$ 0.2 & 140 $\pm$ 1  & 155 $\pm$ 2  \\
C & 132.7 $\pm$ 0.2 & 139 $\pm$ 1  & 151 $\pm$ 2   \\
\hline
\end{tabular}
\end{ruledtabular}
\end{table}

The paraconductivity is generally described by two contributions: $\Delta \sigma = \Delta \sigma_{AL} + \Delta \sigma_{MT}$. The first, in Aslamasov Larkin (AL) framework \cite{asl}, the excess conductivity $\Delta \sigma$ above $T_{c}$ is derived using a microscopic approach by mean field theory, which is considered direct contribution to paraconductivity \cite{wan,ramalho} given by
\begin{equation} \label{eq:al1}
\Delta \sigma_{AL} (\epsilon)= C\epsilon^{-\alpha}
\end{equation}
with $\epsilon = (T - T_{c})/T_{c}$ and
\begin{equation} \label{eq:al2}
C = \frac{e^{2}}{16\hbar d_{AL}}, \quad \alpha = 1 \quad \textrm{for 2D}
\end{equation} 
\begin{equation}\label{eq:al3}
C = \frac{e^{2}}{32\hbar \xi_{z}(0)}, \quad \alpha = \frac{1}{2} \quad \textrm{for 3D.}
\end{equation} 
Here $\alpha$ is the critical exponent related to the dimension of the fluctuations, $\xi_{z}(0)$ is the zero-coherence length in the z-direction for 3D fluctuations, $e$ is the electronic charge, and $d_{AL}$ is characteristic non-superconductor thickness between two superconductor layers. The second contribution arising from the pair-break interaction, which is known as indirect anomalous Maki-Thompson (MT) contribution \cite{maki,thompson}. Since the MT contribution is negligible in cuprate superconductors \cite{naq}, only the AL contribution will be considered ($\Delta \sigma \cong \Delta \sigma_{AL}$)~\cite{ramalho}. 

The evaluation of the AL conductivity contribution can be extracted from the slope of the $\Delta\sigma_{AL}$ versus $\epsilon$ logarithmic plot. The procedure is to fit, for all the three samples, the linear T-dependent resistivity $\rho_{n}=a + bT$ in the interval 220-270K (see figure 4). For all cases, the excess conductivity $\Delta\sigma_{AL}$ was obtained by subtracting the measured conductivity 1/$\rho(T)$ from the linear extrapolated normal-state conductivity  1/$\rho_{n}(T)$ \cite{bat1}:
\begin{equation}
\Delta\sigma_{AL}= \frac{1}{\rho(T)}-\frac{1}{\rho_{n}(T)}.
\end{equation}
From eq. (\ref{eq:al1}) it can be shown that
\begin{equation}
\ln\frac{\Delta\sigma_{AL}}{\sigma_{270K}} = \ln \Big(\frac{C}{\sigma_{270K}}\Big) - \alpha\ln\epsilon.
\end{equation}

Figure 7 shows the dependence of normalized excess conductivity for sample B in form $\ln[\Delta\sigma_{AL}/\sigma_{270K}]$. The $\xi_{z}(0)$ and thickness $d_{AL}$ can be determined from the AL formula using the linear coefficients of the fit when $\alpha$ = 1/2 and 1 (see table \ref{ALfit}). The $\xi_{z}(0)$ and $d_{AL}$ values are relatively large as compared with texture, single crystal, and grain-aligned samples \cite{wan,han1,silee}. These results can be understood with aid of simple grain model, which will be described in the next paragraph. 
\begin{figure}
\includegraphics[angle=90,width=12cm,height=9cm]{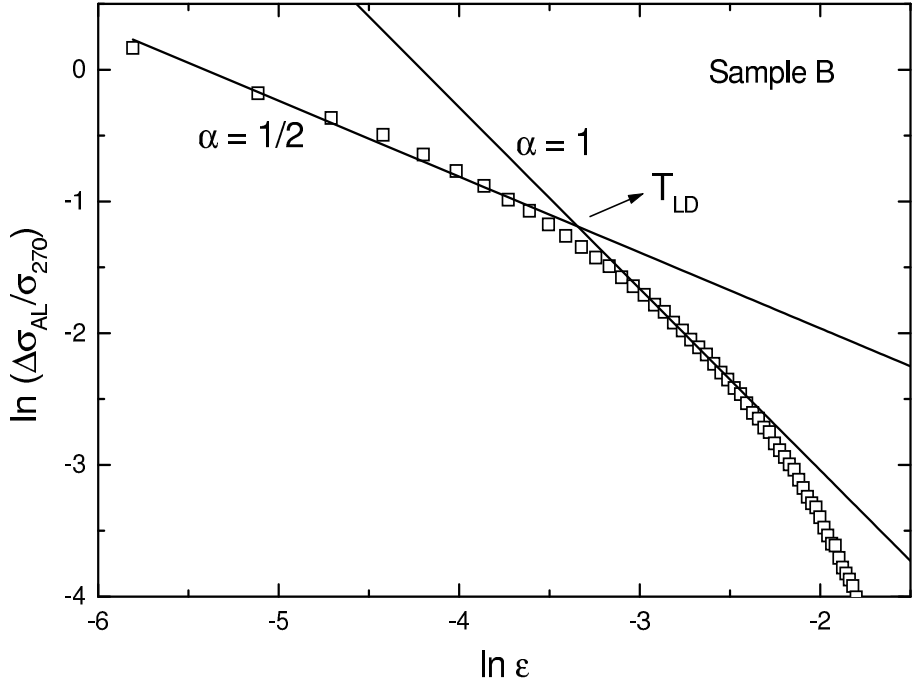}
\caption{Analysis of the excess conductivity normalized in logarithmic scale for the sample B. The linear fitting indicates 3D ($\alpha$ = 1/2) and 2D ($\alpha$ = 1) regimes. The crossover temperature $T_{LD}$ is also indicated by arrow.}
\end{figure}
\begin{table}[h]
\begin{ruledtabular}
\caption{\label{ALfit}Results for the zero-coherent length and distance between planes. The temperature $T_{LD}$ is obtained by intersection between linear fits from analysis of excess conductivity curves.} 
\begin{tabular}{cccccc}
\hline
Sample & $\sigma_{270K}$ $(\Omega~cm)^{-1}$ & $\xi_{z}(0)$ (\AA) & $d_{AL}$ (\AA) & $T_{LD}$ (K)& $d_{LD}$ (\AA)  \\
\hline 
A			 & 10234	& 150	& 1460			& 143 $\pm$ 1 & 287\\
B			 & 29749  & 60	& 1290			& 140 $\pm$ 1 & 113\\
C			 & 18833  & 80  & 1090			& 139 $\pm$ 1 & 150\\
\hline
\end{tabular}
\end{ruledtabular}
\end{table}
Ceramic samples exhibit complex transport behavior because it is composed by particles, where there are grains with pores, microcracks and stacking faults. Intergrain junctions establish the link between different particles, as can be seen in the figure 2. We define a single crystal as a region where discordances are $\lesssim 5^{\circ}$ and grain as being a set of single crystals. As example, a tilt boundary, which is formed from linear sequence of edge dislocations (grain), is shown in figure 8a. Each lamella in the figure 8a represents a grain, and the grains together form agglomeration that is named by particle. 
For instance, the figure 8b shows up a particle formed by 6 grains (lamellas) which presents average thickness of $\cong 4500$ \AA, although there are particles with 6 or 7 lamellas. Moreover, the image contrast of one grain allowed us to identify 3 or 4 different tilts. Therefore, in general, we have observed that there are 3 or 4 single crystals inside each grain, and as consequence, we have estimated that one single crystal has an average size of 1500 \AA. 
\begin{figure}
\includegraphics[width=12cm,height=11cm]{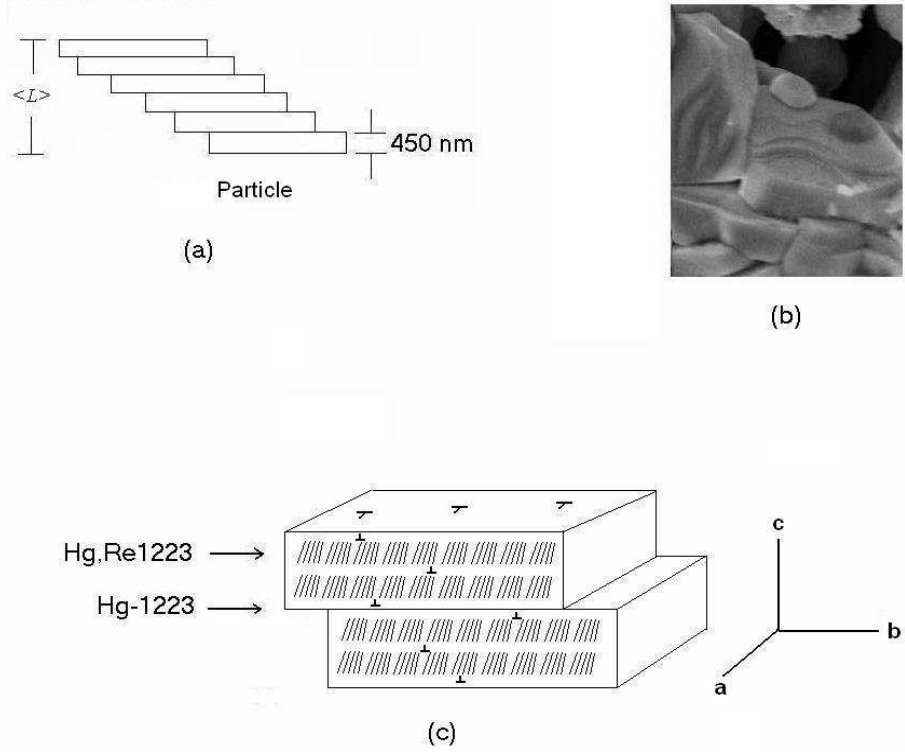}
\caption{The particle can be defined as linear sequence of edge dislocations of the grain growth. In our case, the grain is composed by (Hg,Re)-1223 phase (rich in oxygen) and Hg-1223 phase (poor in oxygen).}
\end{figure}

The analysis of the table \ref{ALfit} suggests that the thickness $d_{Al}$ can be interpreted as an average space between grains. Moreover, this intergrain region is formed by Hg-1223 underdoped crystals, which are in agreement with the values $l$ found in the table \ref{rietveld} for Hg-1223 phase. In principle, this hypothesis can justify our SIS junction type \cite{passos1} and the thickness $d_{Al}$ is lager than the finding in grain-aligned Hg-1223 sample from ref.~\cite{silee}.

\subsubsection{Lawrence-Doniach model}
The grain can be described as proposed in the figure 8c, and in the intragrain region, the Lawrence and Doniach (LD) model \cite{law} can give an appropriate description of the fluctuations. For this model, superconducting layers are coupled by the Josephson effect and the variation of conductivity shows different temperature behavior for different dimensions. 

Following the Schmidt formalism \cite{schmidt}, Lawrence and Doniach derived an expression for the fluctuation-induced in-plane conductivity \cite{law} 
\begin{equation}\label{Tld-1}
\Delta\sigma_{LD} = \frac{e^{2}}{16\hbar d_{LD}} \epsilon^{-1}\Big[1 + \Big(\frac{2\xi_{z}(0)}{d_{LD}}\Big)^{2}\Big]
\end{equation} 
The LD model predicts that near $T_{c}$ a crossover of the dimensionality of the fluctuations occurs and is given by
\begin{equation}\label{Tld-2}
T_{LD}=T_{c}\Big[1 + \Big(\frac{2\xi_{z}(0)}{d_{LD}}\Big)^2\Big].
\end{equation} 
The Lawrence-Doniach approach suggests a change from 2D to 3D behavior at $T_{LD}$ which can be determined from our experimental data by extrapolation of straight lines and taking a crossover point. The change of slope shown in figure 7 indicates a crossover from 2D with $\alpha= 1$ to 3D with $\alpha= 1/2$. Therefore, the resistivity points above $T_{LD}(n)$ in figure 7 are characterized by the 2D exponent $\alpha=1$. The data below $T_{LD}(n)$ are characterized by the 3D exponent $\alpha=1/2$ indicating that in this temperature regime the single crystals are coupled. 

From the Lawrence and Doniach framework one can assume that inside the grain (intragrain region) the non-superconductor region separates the superconducting layers, which is in agreement with the x-ray diffraction analysis where was found two phases: (Hg,Re)-1223 (rich in oxygen) and Hg-1223 (poor in oxygen). The single crystal average size evaluated to Hg-1223 phase, by x-ray diffraction with Rietveld refinement, are in agreement with the thickness obtained by LD approach using $T_{LD}$, $T_{c}$, and $\xi_{z}(0)$ as an input parameters (see table \ref{rietveld} and \ref{ALfit}).

In our simple grain model, $T_{LD}$ physically means that Josephson coupling is taken place between (Hg,Re)-1223 single crystals separated by Hg-1223 phase, in addition, there are indications that above $T_{c}$ the Hg-1223 underdoped phase is an insulating barrier between (Hg,Re)-1223 single crystals, which present fluctuations effects. 

Summarizing, the AL models provides $\xi_{z}(0)$ and $d_{AL}$ parameters which are influenced by intergrain and intragrain regions, however, the Lawrence-Doniach model describes the effects caused from the intragrain fluctuation behavior.

\section{Discussions}
The Rietveld refinement of the XRD measurements have shown that the better spectrum fit was obtained including an extra Hg-1223 phase (poor oxygen) additionally to the main (Hg,Re)-1223 phase (rich at oxygen). The lattice parameter \textit{a} has indicated that the extra Hg-1223 phase is underdoped (poor at oxygen). As intergrain morphologies are similar for all samples (A, B, and C), the second transition in $\chi_{ac}'$ susceptibility is associated with the presence of Hg-1223 phase on grain boundary. As discussed before, from room temperature down to 130 K this Hg-1223 phase is in normal state (insulating phase).

In figure 9 we have presented the  phase diagram $T$ versus $n$, that is, $T_{c}(n)$, $T_{scf}(n)$, $T_{LD}(n)$ and $T^{*}(n)$ for our compounds of the (Hg,Re)-1223 obtained by our resistivity data. The values of $T_{c}(n)$ are in agreement with values of the ac magnetic susceptibility measurement \cite{mtdo}. As discussed above, the values of $T^{*}(n)$ are very settle to be determined. Thus, our results and calculations were made only after a very careful analysis of the voltage-current ($V(I)$) isotherms with the purpose of investigating the best range of current density and temperature where $V(I)$ is in the linear regime for our samples. 
\begin{figure}
\begin{center}
\includegraphics[angle=90,width=12cm,height=9cm]{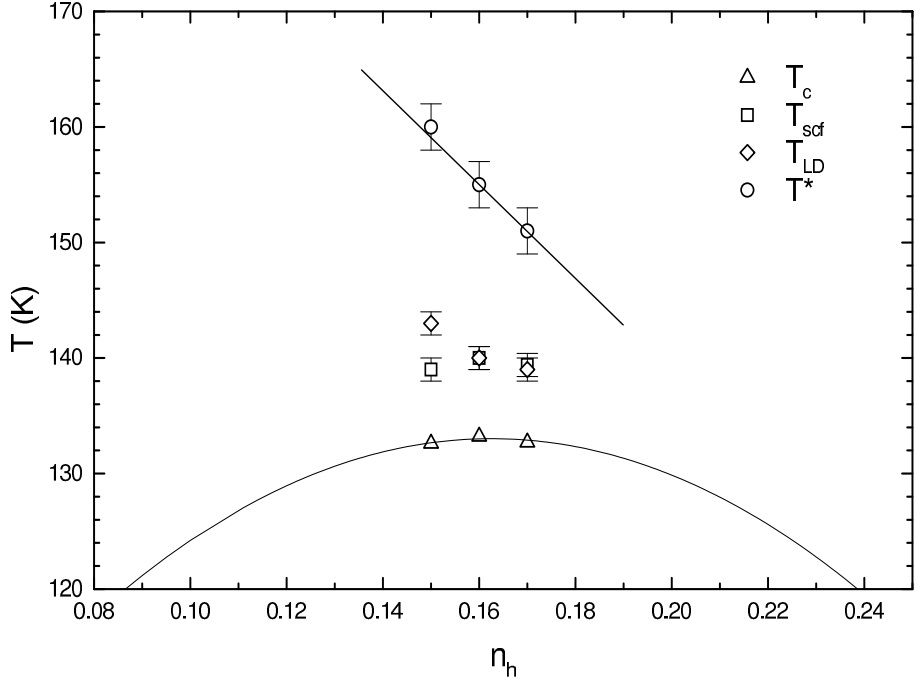}
\caption{The (Hg,Re)-1223 phase diagram with values of $T_{c}$, $T_{scf}$, $T_{LD}$ and $T^{*}$ as function of the charge carrier density. The straight line is drawn as guides to the eye.}
\end{center}
\end{figure}

In order to verify the different nature among pseudogap temperature $T^{*}$, thermodynamic fluctuations temperature $T_{scf}$, and the dimensionality of the fluctuation at a temperature $T_{LD}$, we have performed an investigation on the fluctuation conductivity $\Delta\sigma$. The results indicate that both $T_{LD}(n)$ and $T_{scf}(n)$ are distinct from $T^{*}(n)$.  The $T_{scf}(n)$ curve follows the shape of $T_{c}(n)$ and the difference between $T_{scf}$ and $T_{c}$ is less than 7 K. This is same behavior was also verified by Naqib \textit{et} \textit{al.} \cite{naq} and by Vidal \textit{et} \textit{al.} \cite{vidal}. The low values of $T_{scf}$ with respect to $T^{*}$ does not favor the scenario of the fluctuation of Cooper pairs for the pseudogap phase \cite{eme} as already criticized by Lee \textit{et} \textit{al.} \cite{lee}. 

\section{Conclusion}

We have prepared samples of (Hg,Re) - 1223 superconductors and shown a reliable method to study $T^{*}$ through careful resistivity measurements, as explained above. Our aim is gain some insight about the normal or pseudogap phase of these superconductors. In order to accomplish this task, we have measured $T^{*}(n)$, $T_{LD}(n)$, $T_{scf}(n)$ and $T_{c}(n)$.

From our results and the assumption that the HTSC are inhomogeneous materials \cite{tim,tal,lee}, a possible scenario to normal phase is: $T^{*}$ is the onset of small superconducting islands at the superconductor layers \cite{mel2,mel3}. Initially these islands are isolated and therefore there is a decreasing in the resistivity, as seen by the down turn from the linear regime, but it is still finite. As the temperature decreases, these islands grow and new one appears and there is some overlapping of superconducting regions between the superconducting layers. At $T_{LD}$ the Josephson coupling among the superconducting (intragrain) layers and the system cross from 2D over to 3D behavior. As the temperature goes down the size of the superconducting regions increases. The superconducting regions are large enough to percolate and the sign of this behavior is given by the values of $T_{scf}$ which is just above $T_{c}$~\cite{naq}. 

Another possibility is: $T^{*}$ is the onset of phase separation \cite{lee,mel2,mel3} which is very likely to occur in HTSC. In this case $T^{*}$ has nothing to do with the superconducting phase as proposed by Tallon \textit{et} \textit{al.} \cite{tal} and in agreement with the zinc doped superconductors thin films resistivity measurements \cite{naq}. In this case the superconducting regions start to be formed just above $T_{LD}$ and this temperature marks the onset of coupling among the superconducting layers. At temperatures below $T_{LD}$ the scenario is the same of the above paragraph. However, further studies will be able to distinguish between these two interpretations because the values of $T^{*}$ related to phase segregations are very large, going up to 800 K \cite{tim,tal}, while if the $T^{*}$ is the onset of superconducting islands and it takes much lower values, like those close to the Nernst temperature \cite{yaiu}.

\begin{acknowledgments}
 We would like to thank CNPq Grant CT-Energ 504578/2004-9, CNPq 471536/2004-0, CNPq-FAPERJ Pronex E26/171168/2003, and CAPES for financial supports. Thanks also to Companhia Vale do Rio Doce (CVRD), Companhia Sider\'ugica de Tubar\~ao (CST). We gratefully acknowledge to National Laboratory of Light Synchrotron - LNLS, Brazil.
 \end{acknowledgments}


\end{document}